# Dynamics in Production of Superheavy Nuclei in Low-energy Heavy-ion Collisions*

FENG Zhao-qing, JIN Gen-ming, LI Jun-qing

(*Institute of Modern Physics, Chinese Academy of Sciences, Lanzhou 730000, China*)

**Abstract:** We present a review of the recent progress of theoretical models on the description of the superheavy nucleus formation in heavy ion collisions. Two sorts of reactions that are the fusion-evaporation mechanism and the massive damped collisions to produce superheavy nuclei are discussed. Problems and further improvements of the capture of colliding partners, the formation of compound nucleus and the de-excitation process are pointed out. Possible combinations for the synthesis of the superheavy nuclei in between the products of the cold fusion and $^{48}$Ca induced reactions are proposed by the calculations based on the dinuclear system model and also compared with other approaches. The synthesis of neutron-rich heavy isotopes near sub-shell closure $N=162$ via transfer reactions in the damped collisions of two actinides and the influence of shell closure on the production of heavy isotopes are investigated. Prospective possibility to reach superheavy nuclei near $N=184$ via neutron-rich radioactive beams of high intensity in the future is discussed.

**CLC number:** O572.33　　　　**Document code:** A

## 1 Introduction

The synthesis of heavy or superheavy nuclei is a very important subject in nuclear physics motivated with respect to the island of stability which is predicted theoretically, and has obtained much experimental research with the fusion-evaporation reactions[1—2]. The research of superheavy nuclei attracts much attention in many aspects, e. g., testing the shell model beyond the doubly magic nucleus $^{208}$Pb, exploring the limit of the mass of atomic nucleus and providing a good environment of strong Coulomb field such as quantum electrodynamics(QED) in the super-strong electric field etc..

The existence of the superheavy nucleus (SHN) ($Z \geqslant 106$) is due to strong binding shell effects against the large Coulomb repulsion. However, the shell effects get reduced with increasing the excitation energy of the formed compound nucleus. The synthesis of SHN in experimentally goes back 30 years ago with the multi-nucleon transfer reactions in collisions of two actinide nuclei[3—4]. However, the cross sections of the heavy fragments in strongly damped collisions were found to decrease very rapidly with increasing the atomic number. Combinations with a doubly magic nucleus or nearly magic nucleus are usually chosen owing to the larger reaction $Q$ values. Reactions with $^{208}$Pb or $^{209}$Bi targets were first proposed by Oganessian et al. to synthesize SHN[5]. Neutron-deficient SHN with charged numbers $Z=107—112$ were synthesized using cold fusion reactions for the first time and investigated at GSI (Darmstadt, Germany) with the heavy-ion accelerator UNILAC and the SHIP separator[1, 6]. Experiments on the synthesis of element 113 in the $^{70}$Zn+$^{209}$Bi reaction have been performed successfully at RIKEN (Tokyo, Japan)[7]. However, it is difficulty to produce heavier SHN in the cold fusion reactions because of the smaller production cross sections that are lower than 1 pb for $Z>113$. The superheavy elements $Z=113—118$ were assigned at the Flerov Laboratory of Nuclear Reactions (FLNR) in Dubna (Russia) with the double magic nucleus $^{48}$Ca bombarding actinide nuclei[8−11], in which more neutron-rich SHN were produced and identified by the sequential α decay, unfortunately the decay chain to spontaneous fission of unknown nuclides. Recently, superheavy elements $Z=112$ and $Z=114$ has been further independently verified at GSI and Berkeley (USA) using the reactions $^{48}$Ca+$^{238}$U and $^{48}$Ca+$^{242}$Pu, respectively[12−13]. New heavy isotopes $^{259}$Db and $^{265}$Bh have also been synthesized at HIRFL in Lanzhou (China)[14]. A blank spot exists between the nuclei produced the cold fusion and the $^{48}$Ca induced reactions. Further experimental works are necessary to fill the region and examine the influence of the shell effect in the production of heavy or superheavy isotopes. For that the fusion-evaporation reactions with neutron-rich radioactive ion beams bombarding actinide targets and with the multi-nucleon transfer reactions in collisions of two actinides might be used to fill the region. With the establishment of new facility in the world such as RIBF (RIKEN, Japan), SPIRAL2 (GANIL in Caen, France) and FRIB (MSU, USA), it would be possible to synthesize neutron-rich SHN with the neutron-rich isotope beams of high intensity. Meanwhile, the search of long-lived SHN in the nature is still going on[15].

A reasonable understanding of the formation dynamics of SHN in the massive fusion and transfer reactions is still a challenge for theory. Up to now, many theoretical models have been established, which are mainly based either on several macroscopical degrees of freedom such as the radial elongation, mass or charge asymmetry coupled to the dynamical deformation and to the dissipation of the relative motion energy and the relative angular momentum, or on microscopical degrees of freedom, such as improved isospin dependent quantum molecular dynamics (ImIQMD) model.

The paper is organized as follows. After give a simple introduction of these models in Section 2, we will show some calculated results of fusion dynamics and transfer reactions in the production of heavy or superheavy isotopes in Section 3. In Section 4 a summary is listed.

## 2　Model Descriptions

With performing the successful experiments in laboratories in the world, theoretical models are being established for understanding the formation of SHN. It is different from the low-energy collisions of light systems, the fusion dynamics in the production of SHN is more complicated because of the quasi-fission reactions before the compound nucleus formation. In accordance with the evolution of two heavy colliding nuclei, the dynamical process of the compound nucleus formation and the decay is usually divided into three stages, namely the capture process of the colliding system to overcome the Coulomb barrier, the formation of the compound nucleus to pass over the inner fusion barrier, and the de-excitation of the excited compound nucleus by neutron emission against fission. The transmission in the capture process depends on the incident energy and relative angular momentum of the incident nuclei, which is the same as that in the fusion of light and medium mass systems. The complete fusion of the heavy system after capture in competition with quasi-fission is very important for the estimation of the SHN production. The concept of the "extra-push" energy explains the fusion of two heavy colliding nuclei in the macroscopic dynamical model[16-17]. At present it is still difficult to make an accurate description of the fusion dynamics. After the capture and the subsequent evolution to form the compound nucleus, the thermal compound nucleus will decay by the emission of light particles and γ rays against fission. The three stages will affect the formation of evaporation residues observed in laboratories. The evolution of the whole process of massive heavy-ion collisions is very complicated at near-barrier energies. So the final cross section in the production of SHN is expressed as a sum over all partial waves with angular momentum $J$ at the centre-of-mass energy $E_{c.m.}$.

$$\sigma_{ER}(E_{c.m.}) = \frac{\pi \hbar^2}{2\mu E_{c.m.}} \sum_{J=0}^{J_{max}} (2J+1) \times T(E_{c.m.}, J) P_{CN}(E_{c.m.}, J) W_{sur}(E_{c.m.}, J). \quad (1)$$

Here, $T(E_{c.m.}, J)$ is the transmission probability of the two colliding nuclei overcoming the Coulomb potential barrier in the entrance channel. $P_{CN}$ is the probability that the system will evolve from a touching configuration into the compound nucleus in competition with the quasi-fission reaction of the composite system. The last term is the survival probability of the thermal compound nucleus, which can be estimated with the statistical evaporation model by considering the competition between neutron evaporation and fission.

Most of the theoretical methods on the formation of SHN have a similar viewpoint in the description of the capture and the de-excitation stages, but there is a different description of the process of the compound nucleus formation. There are mainly two sorts of models, depending on whether the compound nucleus is formed along the radial variable (internuclear distance) or by nucleon transfer in a touching configuration which is usually the minimum position of the interaction potential after capture of the colliding system. Several transport models have been established to understand the fusion mechanism of two heavy colliding nuclei leading to SHN formation, such as the macroscopic dynamical model[16-18], the fluctuation-dissipation model[19], the concept of nucleon collectivization[20], the multi-dimensional Langevin equations[21] and the dinuclear system model[22-24]. Recently, ImIQMD model has been applied to simulate the dynamics of the production of SHN in massive fusion reactions and transfer reactions in collisions of two actinides[25-28]. The time dependent Hartree-Fock(TDHF) approach[29] is also used to investigate the dynamics in collisions of $^{238}$U + $^{238}$U. With these models experimental data can be reproduced to a certain extent, and some new results have been predicted. However, these models differ from each other, and sometimes different physical ideas are used.

## 2.1 Capture in collisions of two heavy nuclei

The first stage in the formation of SHN is that the colliding system penetrates (cold fusion reactions) or overcomes (hot fusion reactions) the interaction potential formed by two colliding partners, which is a complicate process associated with the couplings of several degrees of freedom of colliding nuclei, such as radial motion, nucleon transfer, dynamical deformation, vibration and rotation etc.. The coupled channel models that are well known in the description of heavy-ion fusion reactions for light systems are no longer suitable for the capture process of heavy systems owing to the large level numbers at low excitation energies. So the calculated cross sections in the region of sub-barrier energies are much lower than the experimental data. Reasonable description of the capture of two colliding nuclei is of importance for accurately estimating the SHN cross section, especially for the cold fusion reactions. Some phenomenological approaches[20,23,30] or microscopical dynamical models[25-27] have been applied to describe the capture of two heavy nuclei. In the concept of the bar-rier distribution scheme[31], an asymmetric gaussian barriers is used to calculate the transmission probability. The cross section reads as

$$\sigma_{\text{cap}}(E_{\text{c.m.}}) = \frac{\pi \hbar^2}{2\mu E_{\text{c.m.}}} \sum_{J=0}^{\infty} (2J+1) \, T(E_{\text{c.m.}}, J).$$
(2)

Here the penetration probability is given by

$$T(E_{\text{c.m.}}, J) = \int f(B) \times \frac{1}{1 + \exp\left\{-\frac{2\pi}{\hbar\omega(J)}\left[E_{\text{c.m.}} - B - \frac{\hbar^2}{2\mu R_B^2(J)}J(J+1)\right]\right\}} dB,$$
(3)

where the $\hbar\omega(J)$ is the width of the parabolic form of the interaction potential at the position of Coulomb barrier $R_B$. The barrier distribution function is taken as an asymmetric Gaussian form

$$f(B) = \frac{1}{N}\exp\left[-\left(\frac{B-B_m}{\Delta_1}\right)^2\right]$$

at the condition $B \leqslant B_m$ and

$$f(B) = \frac{1}{N}\exp\left[-\left(\frac{B-B_m}{\Delta_2}\right)^2\right]$$

at $B > B_m$ with the relations $B_m = (B_w + B_s)/2$, $\Delta_2 = (B_w - B_s)/2$ and $\Delta_1 = \Delta_2 - \delta$. $\delta$ is changed to reproduce experimental excitation functions and usually is set to be 2—4 MeV in the calculation. $N$ is the normalization constant. $B_w$ and $B_s$ are the heights of the Coulomb barrier at waist-to-waist orientation and of the minimum barrier with varying the dynamical deformations $\beta_1$ and $\beta_2$ of projectile and target, respectively.

The interaction potential reads as

$$V(\{\alpha\}) = V_C(\{\alpha\}) + V_N(\{\alpha\}) + \frac{1}{2}C_1(\beta_1 - \beta_1^0)^2 + \frac{1}{2}C_2(\beta_2 - \beta_2^0)^2, \quad (4)$$

where the $\{\alpha\}$ denotes the symbol of quantities $r$, $\beta_1$, $\beta_2$, $\theta_1$ and $\theta_2$. $\beta_{1,2}$ and $\beta_{1,2}^0$ are the dynamical and static parameters of quadrupole deformation with sign 1 for projectile and 2 for target, respectively. $\theta_{1,2}$ are the polar angles between the radius vector $r$ and the symmetry axes of statically deformed nuclei. With the same potential in the calculation of the potential energy surface in the DNS model, Coulomb potential $V_C$ is obtained by Wong's formula and the nuclear potential $V_N$ calculated by using the double-folding method based on Skyrme force[24]. The stiffness coefficient is calculated by liquid-drop model as

$$C = 4R_N^2\sigma - \frac{3}{10\pi}\frac{Z^2e^2}{R_N}$$

for only including quadrupole deformation with the surface tension coefficient $\sigma$ and the nucleus radius $R_N$. If supposing the ratio of deformation energy is dependent on the mass number as $C_1(\beta_1 - \beta_1^0)^2/C_2(\beta_2 - \beta_2^0)^2 = A_1/A_2$, the dynamical deformation can be represented by only one parameter $\beta = \beta_1 + \beta_2$. Fig. 1 is the Coulomb barrier as a function of the dynamical deformation. The barriers at the nose-nose collisions and at the saddle point are signed in the figure. One can see that a minimum barrier appears with the dynamical deformation and much low than the static barrier, which enhances the sub-barrier cross section. Shown in Fig. 2 is the transmission probability and capture cross section as a function of angular momentum at different incident energies as labeled in the figure which correspond to the compound excitation energies of 8, 12 and 16 MeV, respectively. We also calculated the capture cross sections for the system $^{48}$Ca+$^{238}$U and compared with the experimental data[32] as shown in Fig. 3. The barrier distribution approach can reproduce the experimental data rather well.

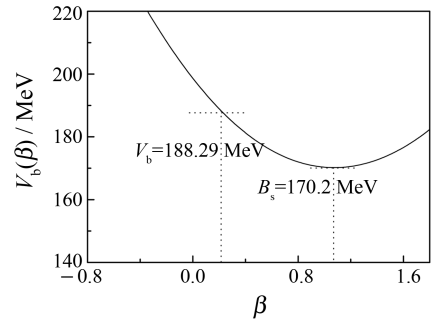

Fig. 1　Dependence of the Coulomb barrier on the dynamical deformation in the reaction $^{48}$Ca+$^{238}$U.

Although the experimental data can be reproduced by the phenomenological barrier distribution approach, the capture cross section in the sub-barrier region is sensitive to the value of the parameter $\Delta_1$ and the dynamics of the capture process in collisions of two heavy nuclides should be studied in detail. For that, we developed the isospin dependent quantum molecular dynamics(IQMD), namely the ImIQMD model[26−27]. The fusion(for light systems) or capture cross section is calculated by

$$\sigma_{cap}(E) = 2\pi \int_0^{b_{max}} b p_{cap}(E, b) db$$
$$= 2\pi \sum_{b=\Delta b}^{b_{max}} b p_{cap}(E, b) \Delta b, \quad (5)$$

where $p_{fus}$ stands for the fusion probability and is given by the ratio of the fusion events $N_{fus}$ to the total events $N_{tot}$. Within the framework of the ImIQMD model, the fusion dynamics or capture dynamics in low-energy heavy-ion collisions has been investigated systematically, such as the static and dynamical interaction potential, dynamical barrier distribution, neck dynamics (the n/p ratio in the neck region, nucleon transfer, neck radius evolution etc.) and fusion or capture excitation functions. Shown in Fig. 4 is a comparison of the calculated static and dynamical interaction poten-tials for head on collisions of the reaction systems $^{48}$Ca+$^{238}$U in the left panel and $^{36}$S+$^{90, 96}$Zr in the right panel. The static interaction potential means that the density distribution of projectile and target is always assumed to be the same as that at initial time, which is a diabatic process and depends on the collision orientations and the mass asymmetry of the reaction systems. For comparison, the proximity results[33] and the adiabatic barrier as mentioned in Ref. [34] are also shown in the figure and the corresponding barrier heights are indicated for the various cases. However, for a realistic heavy-ion collision, the density distribution of the whole system will evolve with the reaction time, which is strongly dependent on the incident energy and impact parameter for a given system. In the calculation of the dynamical potentials, we only pay attention to the fusion events, which give the fusion dynamical barrier. At the same time, stochastic rotation is performed for different simulation events. One can see that the heights of the dynamical barriers are reduced gradually with decreasing the incident energy and increasing the neutron number of the target nucleus. The lowering of the dynamical fusion barrier is in favor of the enhancement of the sub-barrier cross sections, which gives the same concept with the phenomenological barrier distribution approach that is used in DNS calculations[23—24].

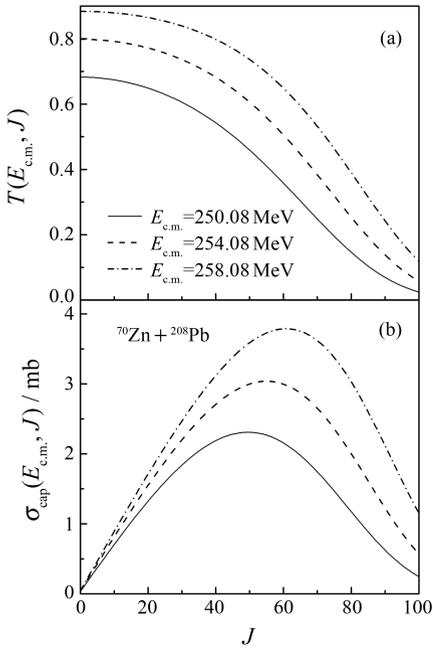

Fig. 2　Calculated partial transmission probability and partial capture cross section in the reaction $^{70}$Zn+$^{208}$Pb at different incident energies.

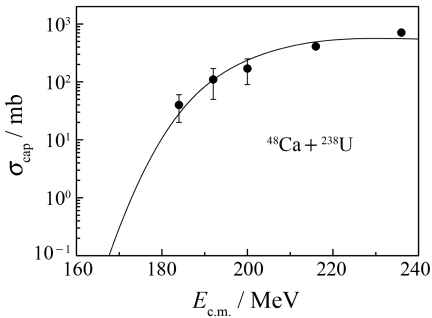

Fig. 3　Comparison of the calculated capture excitation functions with the available experimental data.

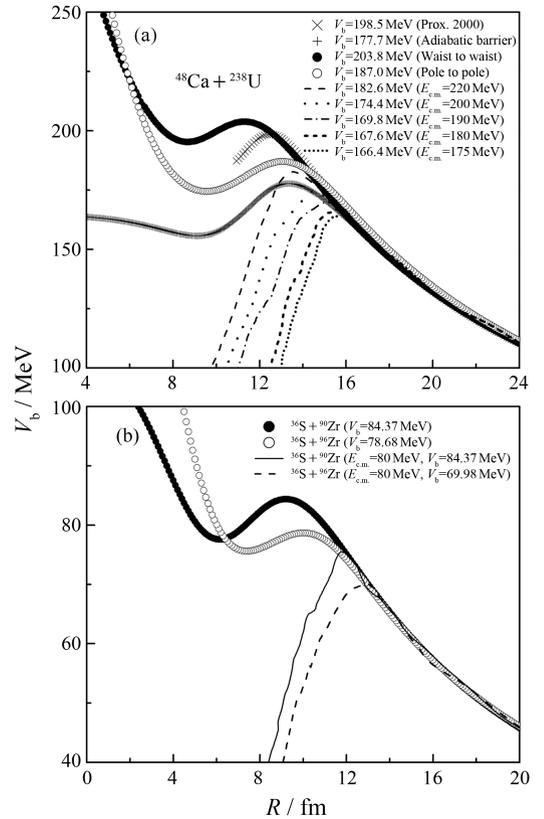

Fig. 4　Comparison of the static fusion barriers (including pole-to-pole and waist-to-waist collisions), adiabatic barrier, proximity results and dynamical barriers at different incident c. m. energies for the reaction $^{48}$Ca +$^{238}$U (a), as well as system dependence of the static and dynamical fusion barriers(b).

To explore more information on the fusion dynamics, we also investigate the distribution of the dynamical fusion barrier, which counts the dynamical barrier per fusion event and satisfies the condition $\int f(B_{fus}) dB_{fus} = 1$. Fig. 5 shows the barrier distribution for head on collisions of $^{58}$Ni+$^{58}$Ni at the center of mass incident energies 96 and 100 MeV, respectively, which correspond to below and above the static barrier $V_b = 97.32$ MeV and a comparison with the system $^{64}$Ni+$^{64}$Ni. The distribution trends towards the low-barrier region with decreasing the incident energy, which can be explained by the slow evolution of the colliding system. The system has enough time to exchange and reorganize nucleons of the reaction partners at lower incident energies. A number of fusion events are located at the sub-barrier region, which is favorable to enhance sub-barrier fusion cross sections. There is a little distribution probability that the fusion barrier is higher than the incident energy 96 MeV owing to dynamical evolution of two touching nuclei. We should note that the fusion events decrease dramatically with incident energy in the sub-barrier region. Neutron-richer system has the distribution towards the low-barrier region owing to the lower dynamical fusion barrier, which favors the enhancement of the fusion cross section. In Fig. 6 we present a comparison of the calculated capture cross sections and the available experimental data for a series of reaction systems. One can see that the calculated results are in good agreement with the experimental data.

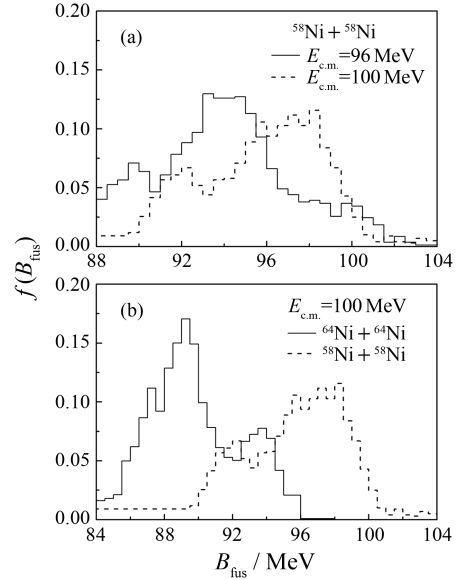

Fig. 5 (a) Distribution of the dynamical fusion barriers at incident energies 96 and 100 MeV in the center of mass frame and (b) comparison of the systems $^{58}$Ni+$^{58}$Ni and $^{64}$Ni+$^{64}$Ni.

## 2.2 Fusion in competition with quasi-fission reaction

The fusion in competition with quasi-fission reaction after capture of colliding partners by the interaction potential is a complicate process associated with evolution of multi-dimensional configurations. The collision orientation and the structure effect of colliding nuclei have a strong influence on the final production of compound nucleus. The dissipation of the angular momentum and the relative motion of kinetic energy is coupled to nucleon transfer in the evolution of the shape from a touching configuration to spherical compound nucleus. The fusion cross section is given by

$$\sigma_{\text{fus}}(E_{\text{c.m.}}) = \frac{\pi \hbar^2}{2\mu E_{\text{c.m.}}} \sum_{J=0}^{\infty} (2J+1)\ T(E_{\text{c.m.}}, J) \times$$

$$P_{\text{CN}}(E_{\text{c.m.}}, J). \quad (6)$$

Theoretical models associated with several degrees of freedom such as mass (charge) asymmetry, radial elongation, neck evolution and deformation dynamics etc., are based on different ideas to get the fusion probability $P_{\text{CN}}$. We will give a simple introduction of these models for describing the formation of SHN in massive fusion reactions.

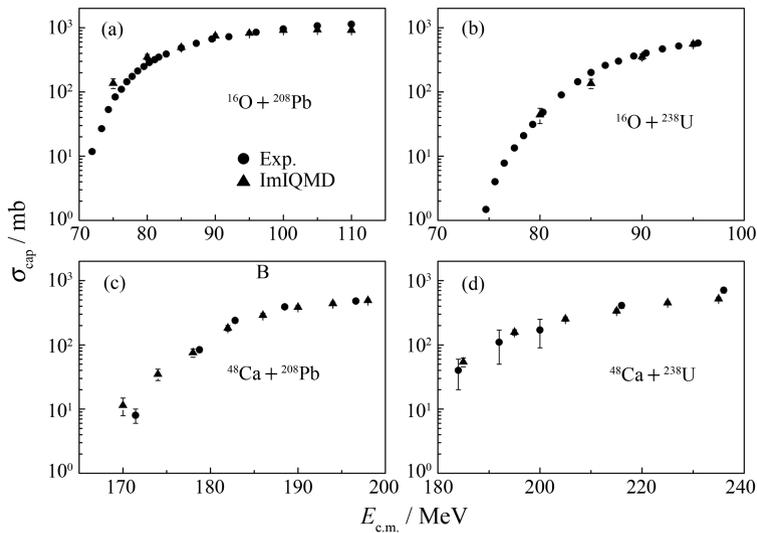

Fig. 6　Comparison of the calculated capture excitation functions and experimental data for the selected reaction systems.

## 2.2.1　Macroscopical dynamics model

The macroscopical dynamical model is the first approach for understanding the formation of SHN proposed by W. J. Swiatecki and his co-workers[16-18], in which the nucleus is assumed to be like a sticking liquid-drop and the fusion or re-separation of two colliding nuclei is a pure dynamics based on the scheme of 'chaotic regime dynamics' (liquid-drop potential energy plus one-body dissipation) at frozen mass asymmetry. Classical equations of motion are used to describe the radial and angular momentum dissipation for two colliding nuclei as follows：

$$\mu \frac{\partial^2 R}{\partial t^2} = -\frac{\partial V_l(R)}{\partial R} - C_R g(R) \frac{\partial R}{\partial t}, \quad (7)$$

$$\frac{\partial l}{\partial t} = -\frac{C_l}{\mu} g(R)(l - l_{\text{st}}). \quad (8)$$

Here $\mu$, $l$ and $l_{\text{st}}$ are the reduced mass, relative angular momentum and sticking angular momentum, respectively. $C_R$ and $C_l$ are the radial and tangential friction coefficients and $V_l(R)$ for the nucleus-nucleus potential.

Within the framework of this model, a colliding system faces with three hurdles for fusing together, namely, the touching configuration, the conditional saddle-point configuration and the unconditional saddle-point configuration. The ideas of the 'extra push' and 'extra-extra push' energies are introduced in the model, which are extra energies to carry the system from the first to second hurdle and from the first to the third hurdle, respectively. The elastic or quasi-elastic scattering takes place when the trajectories does not overcome the threshold at the touching configuration. The configuration is usually close the top of the one-dimensional interaction potential energy of two ap-proaching nuclei. A dinucleus shape is formed in the conditional saddle-point configuration. The energy of conditional equilibrium is only stationary on the condition that the mass asymmetry is fixed (smallness of the neck). When the constriction is not severe, the configuration loses its physical significance. The deep inelastic scattering corresponds to the trajectories that overcome the contact configuration but not the last two hurdles. The unconditional saddle-point configuration is like the fission saddle-point shape and a mononucleus is formed. The associated fission barrier ensures the existence of compound nucleus against disintegration. For a system with large electric charge and angular momentum, the fission barrier disappears and the compound nucleus does not exist. The trajectories that overcome the former hurdles, but not the final one leads to the fast fission reactions. The macroscopical dynamical model successfully explains the suppression of SHN production. Problems are that the shell effect, nucleon transfer, orientation effect, and fluctuation etc. in the formation of compound nucleus do not include in the model.

### 2.2.2 Fluctuation-dissipation model

Two main improvements were performed in the fluctuation-dissipation model based on the macroscopical dynamical model, namely the fluctuation in the collision process and the shell effect in the potential energy. The time evolution of the collective degrees of freedom is given by the following Langevin equation as

$$m \frac{d^2 q}{dt^2} = -\frac{\partial V(q, t)}{\partial q} - \gamma \frac{dq}{dt} + g\Gamma(t), \quad (9)$$

where $m$ and $\gamma$ are the inertial mass and friction coefficient. The last term denotes the Langevin force varying randomly with time, which has the relation by the fluctuation-dissipation theorem as

$$\langle \Gamma(t)\Gamma(t') \rangle = 2\delta(t-t') \quad (10)$$

with $g = \sqrt{\gamma T}$, $T$ is the temperature of the system. The potential energy includes the shell cor-rection and time dependence. In the limit of strong dissipation ($\gamma \to \infty$), the above equation becomes[35]

$$\frac{dq}{dt} = -\frac{1}{\gamma} \frac{\partial V(q, t)}{\partial q} + \frac{g}{\gamma}\Gamma(t). \quad (11)$$

The corresponding distribution probability $P(q, l; t)$ integrating over momentum is given by Smoluchowski equation

$$\frac{\partial P(q, l; t)}{\partial t} = \frac{\partial}{\partial q}\left(\frac{1}{\gamma} \frac{\partial V(q, l; t)}{\partial q} + \frac{T}{\gamma} \frac{\partial}{\partial q}\right) P(q, l; t). \quad (12)$$

The evaporation residue cross section is defined as the probability integrating over the left side at the fission barrier in the final stage of the cooling process[19]. The probability of SHN production is given by

$$P_{eva}(l, t) = \sum_{-\infty}^{q_{sad}} P(q, l; t)dq, \quad (13)$$

where $q_{sad}$ represents the position of the first saddle point. The evaporation residue cross section is calculated by

$$\sigma_{ER}(E_{c.m.}) = \frac{\pi \hbar^2}{2\mu E_{c.m.}} \sum_{l=0}^{l_{max}} (2l+1) \times P_{eva}(l, t=\infty). \quad (14)$$

The fluctuation-dissipation model has been applied to investigate the production cross section of SHN, quasi-fission yields, reaction mechanism of producing SHN etc..

With the concept of two-step model, the sticking and formation probabilities of the two colliding partners are obtained by solving a set of Langevin equations[36]. The shell effect in the fusion dynamics is considered in the model. Calculated results of the fusion cross sections and evaporation residues are consistent with experimental data.

### 2.2.3 Nucleon collectivization model

A certain number of shared nucleons were assumed in the nucleon collectivization model after the colliding system overcomes the Coulomb barri-er and gets in contact. These nucleons move within the whole volume occupied by the nuclear system and form something similar to a neck. In the dynamical evolution, the number of such collectivized nucleons increases whereas the number of nucleons belonging to each individual nucleus decreases[20]. The compound nucleus is formed when the common nucleons expand into the whole system. The inverse process of nucleon de-collectivization leads to the fission channel of heavy nucleus. The experimental evaporation residue cross sections were reproduced rather well by the nucleon collectivization model. However, the shell effect, collision orientation and dynamical deformation are not included in the model.

### 2.2.4 Multidimensional Langevin equations

A unified approach was proposed by Zagrebaev and Greiner, in which the deep inelastic scattering, quasi-fission and fusion-fission were treated by a set of coupled Langevin equations[21]. Seven macroscopic degrees of freedom coupled to the relative motion of radial momentum and the dissipation of angular momentum are included in the model. Influence of dynamical evolution on the intermediate and final products are analyzed and a number of calculations are performed by the model, such as the angular, kinetic energy and charge distributions of the final fragments, the charge and mass distributions of TKE, primary and survived cross sections of fragments etc.. The evaporation residue cross sections were calculated and experimental can be well reproduced. Larger production cross sections of neutron-rich heavy isotopes in multinucleon transfer reactions near sub-shell closure were pointed out by calculations within the model[37].

### 2.2.5 Fusion by diffusion model

The fusion process to form a compound nucleus after overcoming the Coulomb barrier is treated as the diffusion of one-dimensional Brownian motion in a viscous fluid with a repulsive parabolic potential $V(L) = -b(L - L_{max})^2/2$ by elongation $L$[30]. The Smoluchowski equation is used to describe the evolution of the distribution probability

$$G\frac{\partial P(L, t)}{\partial t} = -\frac{\partial (bLP(L, t))}{\partial L} + T\frac{\partial^2 P(L, t)}{\partial L^2}. \quad (15)$$

Here $G$ is the viscosity coefficient of the fluid. A delta function distribution is assumed at the injection point $L_{inj}$ and $P(L, t)$ is a monotonically swelling Gaussian distribution with the time evolution. The portion of the Gaussian distribution overcomes the barrier as $t$ tends to infinity and the probability is given by

$$P(\text{diffuse}) = \frac{1}{2}(1 - \text{erf}\sqrt{H/T}),$$
$$L_{inj} \geqslant L_{max} \quad (16)$$

$$P(\text{diffuse}) = \frac{1}{2}(1 + \text{erf}\sqrt{H/T}),$$
$$L_{inj} < L_{max} \quad (17)$$

Here $H$ is the height of the barrier opposing fusion along the asymmetric fission valley.

Based on the concept of the diffusion process, Liu Zuhua and Bao Jingdong made an improvement that included the neutron flow in the early stage of the evolution[38] and the diffusion was described by a two-variable Smoluchowski equation[39—41]

$$\frac{\partial P(x, y, t)}{\partial t} = [L_x(x, y) + \gamma L_y(x, y)]P(x, y, t). \quad (18)$$

The quantities $L_x$ and $L_y$ are given by

$$L_x(x, y) = -\frac{\partial D_x(x, y)}{\partial x} + D_{xx}\frac{\partial^2}{\partial x^2}, \quad (19)$$

$$L_y(x, y) = -\frac{\partial D_y(x, y)}{\partial y} + D_{yy}\frac{\partial^2}{\partial y^2}. \quad (20)$$

Here $y$ represents the neutron number of light nucleus. A number of calculations and predictions of SHN were performed by the approach. The influence of shell effect and collision orientation on the fusion dynamics is not included in the model.

### 2.2.6 Dinuclear system model

The dinuclear system (DNS)[42] is a molecular configuration of two touching nuclei which keep their own individuality. Such a system has an evolution along two main degrees of freedom: (1) the relative motion of the nuclei in the interaction potential to form a DNS and the decay of the DNS (quasi-fission process) along the $R$ degree of freedom (internuclear motion); (2) the transfer of nucleons in the mass asymmetry coordinate $\eta=(A_1-A_2)/(A_1+A_2)$ between two nuclei, which is a diffusion process of the excited systems leading to the compound nucleus formation. Off-diagonal diffusion in the surface $(A_1, R)$ is not considered since we assume the DNS is formed at the minimum position of the interaction potential of two colliding nuclei. The neck evolution is set to be frozen in the DNS model.

Within the concept of the DNS, the fusion probability was also calculated by using the multi-dimensional Kramers-type expression to get the fusion and quasifission rate by Adamian et al.[22]. In order to describe the fusion dynamics as a diffusion process along proton and neutron degrees of freedom, the fusion probability is obtained by solving a set of master equations numerically in the potential energy surface of the DNS. The time evolution of the distribution probability function $P(Z_1, N_1, E_1, t)$ for fragment 1 with proton number $Z_1$ and neutron number $N_1$ and with excitation energy $E_1$ is described by the following master equations,

$$\frac{d P(Z_1, N_1, E_1, t)}{dt} = \sum_{Z_1'} W_{Z_1, N_1; Z_1', N_1}(t)[d_{Z_1, N_1} P(Z_1', N_1, E_1', t) - d_{Z_1', N_1} P(Z_1, N_1, E_1, t)] + \sum_{N_1'} W_{Z_1, N_1; Z_1, N_1'}(t) \times [d_{Z_1, N_1} P(Z_1, N_1', E_1', t) - d_{Z_1, N_1'} P(Z_1, N_1, E_1, t)] - [\Lambda_{qf}(\Theta(t)) + \Lambda_{fis}(\Theta(t))] P(Z_1, N_1, E_1, t). \quad (21)$$

Here $W_{Z_1, N_1; Z_1', N_1}$ ($W_{Z_1, N_1; Z_1, N_1'}$) is the mean transition probability from the channel $(Z_1, N_1, E_1)$ to $(Z_1', N_1, E_1')$ (or $(Z_1, N_1, E_1)$ to $(Z_1, N_1', E_1')$), and $d_{Z_1, N_1}$ denotes the microscopic dimension corresponding to the macroscopic state $(Z_1, N_1, E_1)$. The sum is taken over all possible proton and neutron numbers that fragment $Z_1'$, $N_1'$ may take, but only one nucleon transfer is considered in the model with the relation $Z_1'=Z_1\pm 1$ and $N_1'=N_1\pm 1$. The excitation energy $E_1$ is determined by the dissipation energy from the relative motion and the potential energy surface of the DNS. The motion of nucleons in the interacting potential is governed by the single-particle Hamiltonian[23-24]. The evolution of the DNS along the variable $R$ leads to the quasi-fission of the DNS. The quasi-fission rate $\Lambda_{qf}$ and the fission rate $\Lambda_{fis}$ of heavy fragment are estimated with the one-dimensional Kramers formula[23, 43]. The single-particle Hamiltonian is given by[23, 44]

$$H(t) = H_0(t) + V(t) \quad (22)$$

with

$$H_0(t) = \sum_K \sum_{\nu_K} \varepsilon_{\nu_K}(t) a^\dagger_{\nu_K}(t) a_{\nu_K}(t),$$
$$V(t) = \sum_{K, K'} \sum_{\alpha_K, \beta_{K'}} u_{\alpha_K, \beta_{K'}}(t) a^\dagger_{\alpha_K}(t) a_{\beta_{K'}}(t)$$
$$= \sum_{K, K'} V_{K, K'}(t). \quad (23)$$

Here the indices $K$, $K'$ ($K, K'=1, 2$) denote the fragments 1 and 2. The quantities $\varepsilon_{\nu_K}$ and $u_{\alpha_K, \beta_{K'}}$ represent the single particle energies and the interaction matrix elements, respectively. The single particle states are defined with respect to the centers of the interacting nuclei and are assumed to be orthogonalized in the overlap region. So the annihilation and creation operators are dependent on time. The single particle matrix elements are parameterized by

$$u_{\alpha_K, \beta_{K'}}(t) = U_{K, K'}(t) \times \left\{ \exp\left[-\frac{1}{2}\left(\frac{\varepsilon_{\alpha_K}(t)-\varepsilon_{\beta_{K'}}(t)}{\Delta_{K, K'}(t)}\right)^2\right] - \delta_{\alpha_K, \beta_{K'}} \right\}, \quad (24)$$

which contains some parameters $U_{K, K'}(t)$ and $\Delta_{K, K'}(t)$. The detailed calculation of these parameters and the mean transition probabilities were described in Refs. [23, 44]. The evolution of the DNS along the variable $R$ leads to the quasi-fission of the DNS.

In the relaxation process of the relative motion, the DNS will be excited due to the dissipation of the relative kinetic energy. The excited system opens a valence space $\Delta_{\varepsilon_K}$ in fragment $K$ ($K=1$, 2), which has a symmetrical distribution around the Fermi surface. Only the particles in the states within the valence space are actively involved in excitation and transfer. The averages on these quantities are performed in the valence space:

$$\Delta_{\varepsilon_K} = \sqrt{\frac{4\varepsilon_K^*}{g_K}}, \quad \varepsilon_K^* = \varepsilon^* \frac{A_K}{A}, \quad g_K = \frac{A_K}{12}, \quad (25)$$

where $\varepsilon^*$ is the local excitation energy of the DNS, which provides the excitation energy for the mean transition probability. There are $N_K = g_K \Delta_{\varepsilon_K}$ valence states and $m_K = N_K/2$ valence nucleons in the valence space $\Delta_{\varepsilon_K}$, which give the dimension

$$d(m_1, m_2) = \binom{N_1}{m_1}\binom{N_2}{m_2}.$$

The local excitation energy is defined as

$$\varepsilon^* = E_x - (U(A_1, A_2) - U(A_P, A_T)). \quad (26)$$

Here $U(A_1, A_2)$ and $U(A_P, A_T)$ are the driving potentials of fragments $A_1$, $A_2$ and fragments $A_P$, $A_T$ (at the entrance point of the DNS), respectively. The excitation energy $E_x$ of the composite system is converted from the relative kinetic energy loss, which is related to the Coulomb barrier $B$[24] and determined for each initial relative angular momentum $J$ by the parametrization method of the classical deflection function[45-46]. So $E_x$ is coupled with the relative angular momentum. The dissipation of the relative angular momentum and the relative motion of kinetic energy is given in detail in Ref. [47].

The local excitation energy is determined by the excitation energy of the composite system and the potential energy surface of the DNS. The potential energy surface(PES) of the DNS is given by

$$U(\{\alpha\}) = B(Z_1, N_1) + B(Z_2, N_2) - [B(Z, N) + V_{\text{rot}}^{\text{CN}}(J)] + V(\{\alpha\}) \quad (27)$$

with $Z_1 + Z_2 = Z$ and $N_1 + N_2 = N$. Here the symbol $\{\alpha\}$ denotes the sign of the quantities $Z_1$, $N_1$, $Z_2$, $N_2$; $J$, $R$; $\beta_1^0$, $\beta_2^0$, $\theta_1$, $\theta_2$. The $B(Z_i, N_i)$ ($i=1$, 2) and $B(Z, N)$ are the negative binding energies of the fragment ($Z_i$, $N_i$) and the compound nucleus ($Z$, $N$), respectively, which are calculated from the liquid drop model, in which the shell and the pairing corrections are included reasonably. $V_{\text{rot}}^{\text{CN}}$ is the rotation energy of the compound nucleus. $\beta_i^0$ represent the quadrupole deformations of the two fragments. $\theta_i$ denote the angles between the collision orientations and the symmetry axes of deformed nuclei. The interaction potential between fragment ($Z_1$, $N_1$) and ($Z_2$, $N_2$) includes the nuclear, Coulomb and centrifugal parts, the details are given in Ref. [24]. In the calculation, the distance $R$ between the centers of the two fragments is chosen to be the value which gives the minimum of the interaction potential, in which the DNS is considered to be formed. So the PES depends on the proton and neutron numbers of the fragment. In Fig. 7 we give the potential energy surface in the reaction $^{30}$Si + $^{252}$Cf as functions of protons and neutrons of the fragments in the left panel (2D PES). The incident point is shown by the open circle and the minimum trajectory in the PES is added by the white line. The driving potential as a function of the mass asymmetry of two fragments $\eta = (A_1 - A_2)/(A_1 + A_2)$ (1D PES) that was calculated in Refs. [23-24, 48] is given in the right panel and also compared with the minimum trajectory of the 2D PES shown in the left panel. In the 1D PES, we chose the way which gives the minimum value of the PES after transferring proton or neutron from the incident point. So the 1D PES only depends on the mass asymmetry of two fragments (one degree of freedom). For the system $^{30}$Si + $^{252}$Cf, the driving potential at the incident point in 1D PES is located at the position of the maximum value, so there is no inner fusion barrier, which results in too large fusion probability. Therefore, we solve the master equations within the 2D PES in order to correctly get the fusion probability for those systems with larger entrance mass asymmetries and quadrupole deformations (2D master equations)[49]. The master equations can be also solved within the 1D PES (1D master equations). Both methods give the same values for the systems with the smaller projectile-target mass asymmetries and also smaller quadrupole deformations of the initial combinations, such as the cold fusion reactions.

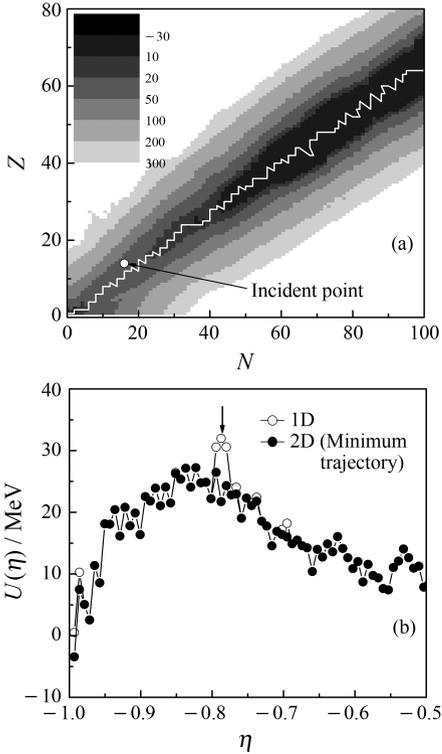

Fig. 7 The potential energy surface of the DNS in the reaction $^{30}$Si+$^{252}$Cf as functions of the protons and neutrons of the fragments (a) and the mass asymmetry coordinate (b).

The formation probability of the compound nucleus at the Coulomb barrier $B$ and for the angular momentum $J$ is given by[23−24]

$$P_{CN}(E_{c.m.}, J, B) = \sum_{Z_1=1}^{Z_{BG}} \sum_{N_1=1}^{N_{BG}} P(Z_1, N_1, E_1, \tau_{int}) .$$
(28)

The interaction time $\tau_{int}$ in the dissipation process of two colliding partners is dependent on the incident energy $E_{c.m.}$ and the quantities $J$ and $B$. We obtain the fusion probability as

$$P_{CN}(E_{c.m.}, J) = \int f(B) P_{CN}(E_{c.m.}, J, B) dB ,$$
(29)

where the barrier distribution function is taken as an asymmetric Gaussian form.

### 2.3 Survival probability of superheavy nucleus

The excited heavy or superheavy compound nucleus formed by fusing two colliding partners in competition with quasi-fission reaction will disintegrate into two parts owing to fission process with larger probability, and more fragments with emitting neutrons, light charged particles and γ rays. Superheavy residues observed experimentally are the yields after the thermal compound nuclei evaporate light particles or γ rays in competition with fission reactions. The branch ratio is determined by the decay width, which can be calculated by the statistical model. If supposing the electric dipole radiation is primary in the γ emission, the evaporation width is written as[50]

$$\Gamma_\gamma(E^*, J) = \frac{3}{\rho(E^*, J)} \times \int_0^{E_\gamma} \rho(E^* - E_{rot} - \varepsilon, J) f_{E1}(\varepsilon) d\varepsilon,$$
(30)

where the $E_\gamma = E^* - E_{rot} - \Delta$ is given by the excitation energy $E^*$, the rotation energy $E_{rot} = J(J+1) \times \hbar^2/2\xi$ with the moment of inertia $\xi = 0.4\lambda MR^2$ and the pairing energy $\Delta = 12\chi/\sqrt{A}$ in MeV ($\chi = -1, 0$ and 1 for odd-odd, odd-even and even-even nuclei, respectively). Here $M$ and $R$ are the mass and radius of the compound nucleus, and the coefficient $\lambda = 0.4$ is the factor for correcting the rigid-body approximation. The strength function is given by

$$f_{E1}(\varepsilon) = \frac{4}{3\pi} \frac{1+\kappa}{mc^2} \frac{e^2}{\hbar c} \frac{NZ}{A} \frac{\Gamma_G \varepsilon^4}{(\Gamma_G \varepsilon)^2 + (E_G^2 - \varepsilon^2)^2}$$
(31)

with $\kappa = 0.75$. Here $m$, $\Gamma_G$ and $E_G$ are the nucleon mass, width and energy of the dipole resonance, respectively. For heavy nucleus[51], they are taken as $\Gamma_G = 5$ MeV and

$$E_G = \frac{167.23}{A^{1/3}\sqrt{1.959 + 14.074A^{-1/3}}} \quad \text{MeV.} \quad (32)$$

The evaporation width of the light particles is usually calculated by using the Weisskopf's evaporation model[52]

$$\Gamma_\nu(E^*, J) = (2s_\nu + 1)\frac{m_\nu}{\pi^2\hbar^2\rho(E^*, J)}\int_0^{E_\nu}\varepsilon \times$$
$$\rho(E^* - B_\nu - \delta_n - E_{rot} - \varepsilon, J)\sigma_{inv}(\varepsilon)d\varepsilon, \quad (33)$$

where $s_\nu$, $m_\nu$ and $B_\nu$ are spin, mass and binding energy of evaporating particle $\nu$ and $E_\nu = E^* - B_\nu - E_{rot} - \Delta - \delta_n$. $\delta_n$ is a factor for revising neutron evaporation, which is taken as $\delta_n = 12/\sqrt{A}$ for odd-neutron and even-proton nucleus, $\delta_n = 0$ for others[53]. The inverse cross section reads as

$$\sigma_{inv}(\varepsilon) = \pi R_\nu^2\left(1 - \frac{V_\nu}{\varepsilon}\right) \quad \text{for } \varepsilon \geq V_\nu$$
$$= 0 \quad \text{for } \varepsilon < V_\nu \quad (34)$$

with

$$R_\nu = 1.21[(A - A_\nu)^{1/3} + A_\nu^{1/3}] + (3.4/\sqrt{\varepsilon})\delta_{\nu,n}, \quad (35)$$

where $A_\nu$ is the mass number of emitting particles $\nu$(n, p, d, α, ...). The Coulomb barrier of charged particles is given by

$$V_\nu = \frac{K_\nu(Z - Z_\nu)Z_\nu}{R_\nu + 1.6} \quad \text{MeV}, \quad (36)$$

and $K_\nu = 1.15$ for proton, $K_\nu = 1.32$ for d and α.

The fission width is usually calculated by Bohr-Wheeler formula[54]

$$\Gamma_f(E^*, J) = \frac{1}{2\pi\rho_f(E^*, J)} \times$$
$$\int_0^{E_f}\frac{\rho_f(E^* - B_f - \delta_f - E_{rot} - \varepsilon, J)d\varepsilon}{1 + \exp[-2\pi(E^* - B_f - E_{rot} - \varepsilon)/\hbar\omega]},$$
$$(37)$$

where the width of potential pocket is taken as $\hbar\omega = 2.2$ MeV and $E_f = E^* - B_f - E_{rot} - \Delta - \delta_f$. $\delta_f$ is a correct factor of the fission barrier $B_f$, which is set to be $\delta_f = \delta$ for even-A nucleus, and $\delta_f = 0$ for others. The fission barrier includes the macroscopical part and the microscopical shell correction as

$$B_f(E^*) = B_f^{LD} - E_{sh}\exp(-\gamma E^*). \quad (38)$$

Here the macroscopical fission barrier is calculated by Liquid-Drop model[55]

$$B_f^{LD} = \begin{cases} 0.38(0.75 - x)E_{s0}, & 1/3 < x < 2/3 \\ 0.83(1 - x)^3 E_{s0}, & 2/3 \leq x < 1 \end{cases}$$
$$(39)$$

with the fissility parameter $x = E_{C0}/2E_{s0}$. Here $E_{C0}$ and $E_{s0}$ are the Coulomb and surface energy of the compound nucleus, respectively, which are given by the well known Myers-Swiatecki mass formula as[56]

$$E_{C0} = 0.7053Z^2/A^{1/3} \quad \text{MeV}, \quad (40)$$

$$E_{s0} = 17.944[1 - 1.7826\eta^2]A^{2/3} \quad \text{MeV} \quad (41)$$

with relative neutron excess $\eta = (N - Z)/A$. The shell correction energy $E_{sh}$ can be get from nuclear mass tables[57]. The level density parameter in the calculation of the fission width is set to be $a_f = 1.1a$ for the level density at saddle point $\rho_f$. Shown in Fig. 8 is the decay width of the fission, light parti-cles and γ rays as a function of the excitation energy of the compound nucleus $^{290}114$ at angular momentum $J = 0$. One can see that γ rays only have contribution at low excitation energy ($E^* < 10$ MeV). The light particles can be emitted at excitation energies larger than their binding energies. Therefore, the competition between fission and neutron evaporation is dominant in the range of SHN synthesis at excitation energies 8 MeV $< E^* < 60$ MeV.

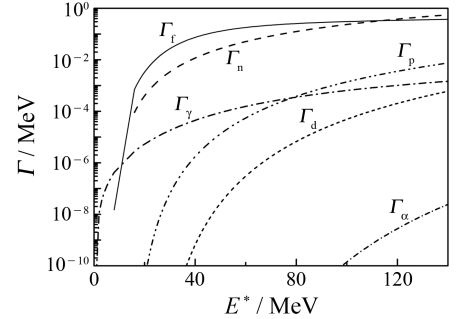

Fig. 8　Excitation energy dependence of the decay width of compound nucleus $^{290}114$ at $J = 0$.

The level density at ground state is expressed by the back-shifted Bethe formula[58] with the spin

cut-off model as

$$\rho(E^*, J) = K_{rot} K_{vib} \frac{2J+1}{24\sqrt{2}\sigma^3} a^{-1/4}(E^* - \Delta)^{-5/4} \times$$
$$\exp[2\sqrt{a(E^* - \Delta)} - \frac{(J+1/2)^2}{2\sigma^2}] , \quad (42)$$

where $K_{rot}$ and $K_{vib}$ are the coefficients of the rotational and vibrational enhancements. The spin cut-off parameter is calculated by the formula:

$$\sigma^2 = \frac{T\zeta_{r.b}}{\hbar^2} , \quad (43)$$

where the rigid-body moment of inertia has a relation $\zeta_{r.b} = 0.4MR^2$ with nuclear mass $M$ and radius $R$. The level density parameter is related to the shell correction energy $E_{sh}(Z, N)$ and the excitation energy $E^*$ of the nucleus as

$$a(E^*, Z, N) = \tilde{a}(A) \times \frac{1 + E_{sh}(Z, N) f(E^* - \Delta)}{(E^* - \Delta)} . \quad (44)$$

Here, $\tilde{a}(A) = \alpha A + \beta A^{2/3} b_s$ is the asymptotic Fermi-gas value of the level density parameter at high excitation energy. The shell damping factor is given by

$$f(E^*) = 1 - \exp(-\gamma E^*) \quad (45)$$

with $\gamma = \tilde{a}/(\varepsilon A^{4/3})$. All of the used parameters in the DNS calculation are listed in Table 1. In Fig. 9 we give the level density parameters of different nuclides at ground state compared them with two empirical formulas $a(A) = A/8$, and $A/12$. One can see that the strong shell effects appear. So the structure effect is clear in the level density, which is very significant in the estimation of the survival probability.

Table 1　Parameters used in the calculation of the level density

| $K_{rot}$ | $K_{vib}$ | $b_s$ | $\alpha$ | $\beta$ | $\varepsilon$ |
|---|---|---|---|---|---|
| 1 | 1 | 1 | 0.114 | 0.098 | 0.4 |

After considering the competition between neutron evaporation and fission process, the survival probability of the excited compound nucleus is expressed as follows:

$$W_{sur}(E^*_{CN}, x, J) = P(E^*_{CN}, x, J) \times \prod_{i=1}^{x} \left( \frac{\Gamma_n(E^*_i, J)}{\Gamma_n(E^*_i, J) + \Gamma_f(E^*_i, J)} \right)_i , \quad (46)$$

where $E^*_{CN}$, $J$ are the excitation energy and the spin of the compound nucleus, respectively. $E^*_i$ is the excitation energy before evaporating the $i$th neutron, which has the relation

$$E^*_{i+1} = E^*_i - B^n_i - 2T_i , \quad (47)$$

with the initial condition $E^*_1 = E^*_{CN}$. $B^n_i$ is the separation energy of the $i$th neutron. The nuclear temperature $T_i$ is given by $E^*_i = aT_i^2 - T_i$ with the level density parameter $a$. $P(E^*_{CN}, x, J)$ is the realization probability of emitting $x$ neutrons and the detailed calculation is shown in Refs [23, 59—60]. We calculated the survival probability in the 1n—5n evaporation channels of the compound nucleus $^{290}114$ at angular momentum $J=0$ as a function of the excitation energy, and also the angular momentum dependence at 40 MeV excitation energy as shown in Fig. 10. The position of the maximal values of each channels is mainly determined by the realization probability. The survival probability decreases with the angular momentum owing to the fissile nucleus. So we take the maximum angular momentum $J_{max} = 30$ to estimate the cross section in Eq. (1) in the fusion-evaporation reactions.

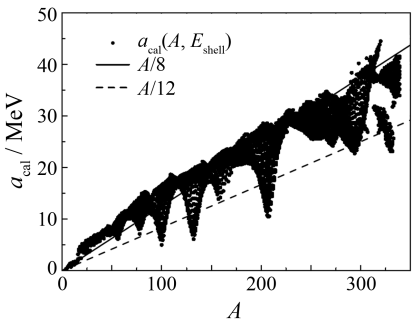

Fig. 9　Calculated level density parameters as a function of the atomic mass and compared with empirical formulas.

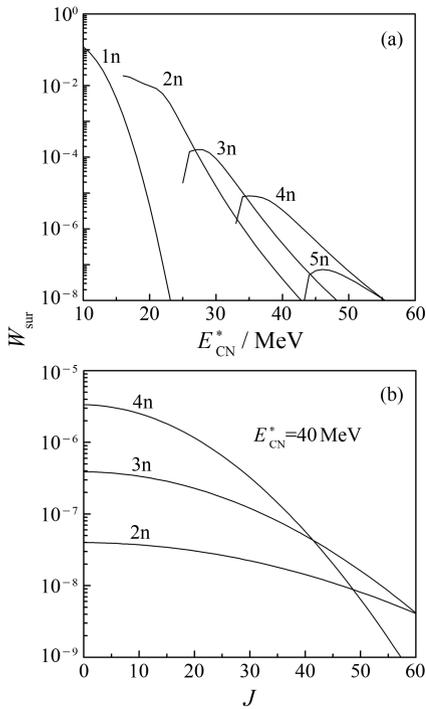

Fig. 10　Survival probability of the compound nucleus $^{290}114$ as functions of angular momentum and excitation energy.

## 3　Results and Discussions

### 3.1　Production cross sections of heavy and superheavy nuclei

The evaporation residues observed in laboratories by the sequential $\alpha$ decay are mainly produced by the complete fusion reactions, in which the fusion dynamics and the structure properties of the compound nucleus affect their production. Within the framework of the DNS model, in Fig. 11 we show the calculated maximal production cross sections of superheavy elements $Z=102—120$ in the cold fusion reactions by evaporating one neutron, in the $^{48}$Ca induced reactions with actinide targets by evaporating three neutrons, and the experimental data[1-2, 6, 61]. The production cross sections decrease rapidly with increasing the charge number of the synthesized compound nucleus in the cold fusion reactions, such as from 0.2 $\mu$b for the reaction $^{48}$Ca+$^{208}$Pb down to 1 pb for $^{70}$Zn+$^{208}$Pb, and even below 0.1 pb for synthesizing $Z\geqslant 113$[24]. It seems to be difficult to synthesize superheavy elements $Z\geqslant 113$ in the cold fusion reactions at the present facilities. The calculated results show that the $^{48}$Ca induced reactions have smaller production cross sections with $^{232}$Th target, but are in favor of synthesizing heavier SHN ($Z\geqslant 113$) because of the larger cross sections[62-63]. The experimental data also give such trends. In the DNS concept, the inner fusion barrier increases with reducing mass asymmetry in the cold fusion reactions, which leads to a decrease of the formation probability of the compound nucleus. However, the $^{48}$Ca induced reactions have no such increase of the inner fusion barrier for synthesizing heavier SHN. Because of the larger transmission and the higher fusion probability, we obtain larger production cross sections for synthesizing SHN ($Z\geqslant 113$) in the $^{48}$Ca induced reactions although these reactions have a smaller survival probability than those in the cold fusion reactions. It is still a good way to synthesize heavier SHN by using the $^{48}$Ca induced reactions. Of course, further experimental data are anticipated to be obtained in the future. However, the actinide targets are difficult to be handled in experiments synthesizing heavier SHN.

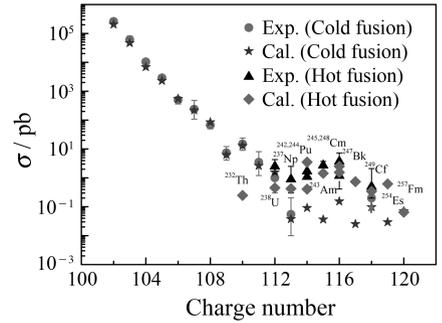

Fig. 11　Maximal production cross sections of superheavy elements $Z=102-120$ in cold fusion reactions based on $^{208}$Pb and $^{209}$Bi targets with projectile nuclei $^{48}$Ca, $^{50}$Ti, $^{54}$Cr, $^{58}$Fe, $^{64}$Ni, $^{70}$Zn, $^{76}$Ge, $^{82}$Se, $^{86}$Kr and $^{88}$Sr, in $^{48}$Ca induced reactions with actinide targets by evaporating 3 neutrons, and in comparison with available experimental data.

Uranium is the heaviest element existing in the nature. It has a larger mass asymmetry con-structed as a target in the fusion reactions with the various neutron-rich light projectiles. The isotope $^{238}$U is the neutron-richest nucleus in the U isotopes and often chosen as the target for synthesizing SHN. In Fig. 12 we draw the evaporation residue excitation functions of the reactions $^{40}$Ar, $^{50}$Ti, $^{54}$Cr, $^{64}$Ni+$^{238}$U in the 2n—5n channels. The results show that the 4n channel in the reaction $^{40}$Ar+$^{238}$U has larger cross sections about 2.1 pb at an excitation energy 42 MeV. The reactions $^{50}$Ti, $^{54}$Cr, $^{64}$Ni+$^{238}$U lead to the cross section smaller than 0.1 pb. Calculations show that the isotopes $^{235}$U and $^{238}$U are favorable in producing SHN. The cross sections are reduced with increasing the mass numbers of the projectiles. Other reaction mechanisms to synthesize SHN have to be investigated with theoretical models, such as the massive transfer reactions, and the complete fusion reactions induced by weakly bound nuclei. Work in these directions is in progress within the framework of the DNS model.

Neutron-deficient SHN with charged numbers

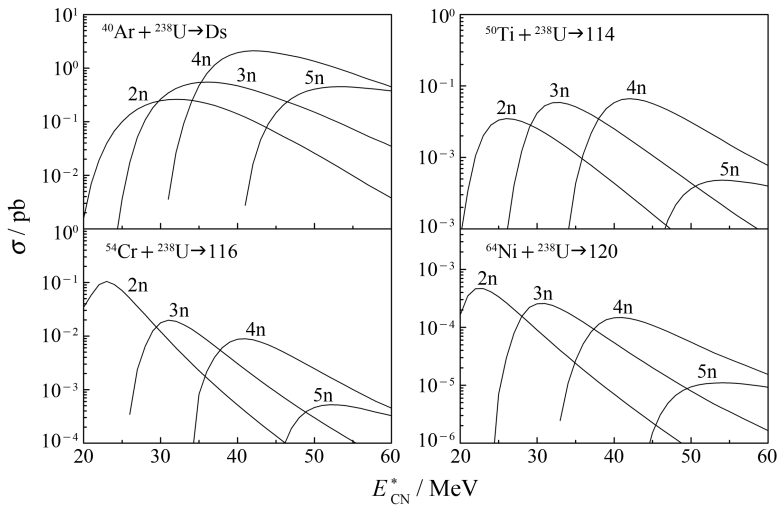

Fig. 12 The evaporation residue excitation functions in the reactions $^{40}$Ar, $^{50}$Ti, $^{54}$Cr, $^{64}$Ni+$^{238}$U.

$Z=107$—113 were synthesized successfully in the cold fusion reactions. The evaporation residues was observed by the consecutive $\alpha$ decay until to take place spontaneous fission of known nuclides, in which the fusion dynamics and the structure properties of the compound nucleus have strongly influence in the production of SHN. Recently more neutron-rich and heavier SHN with charged numbers $Z=113$—118 were produced in the fusion-evaporation reactions of $^{48}$Ca bombarding actinide targets. Superheavy residues were also identified by the consecutive $\alpha$ decay, unfortunately to spontaneous fission of unknown nuclides. Neutron-rich projectile-target combinations are necessary to be chosen so that superheavy residues approach the "island of stability" with the doubly magic shell closure beyond $^{208}$Pb at the position of protons $Z=114$—126 and neutrons $N=184$. New SHN between the isotopes of the cold fusion and the $^{48}$Ca induced reactions are of importance for the structure studies themselves and also as comparable nuclides for identifying heavier SHN in the future. Superheavy element Ds($Z=110$) was successfully synthesized in the cold fusion reactions[1, 64]. The production of the SHN depends on the isotopic combinations of projectiles and targets. For example, the maximal cross section is $3.5^{+2.7}_{-1.8}$ pb for the reaction $^{62}$Ni+$^{208}$Pb→$^{269}$Ds+1n, however $15^{+9}_{-6}$ pb for the reaction $^{64}$Ni+$^{208}$Pb→$^{271}$Ds+1n[64]. In the DNS model, the isotopic trends are mainly determined by both the fusion and survival probabilities. When the neutron number of the projectile is increasing, the DNS gets more symmetrical and the fusion probability decreases if the DNS does not consist of more stable nuclei due to a higher inner fusion barrier. A smaller neutron separation energy and a larger shell correction lead to a larger survival probability. The compound nucleus with closed neutron shells has larger shell correction energy and neutron separation energy. Calculations were performed for the reactions $^{30}$Si+$^{248,250}$Cm, $^{36}$S+$^{244}$Pu and $^{40}$Ar+$^{238}$U to produce superheavy element Ds as shown in Fig. 13. Combination with $^{248}$Cm has the larger cross section in the 4n channel than the isotope $^{250}$Cm due to the larger value of survival probability. The 4n channels in the systems $^{30}$Si+$^{248,250}$Cm and $^{40}$Ar+$^{238}$U and the 3n channel in the reaction $^{30}$Si+$^{248}$Cm are feasible in the synthesis of new isotopes of SHN $^{274-276}$Ds.

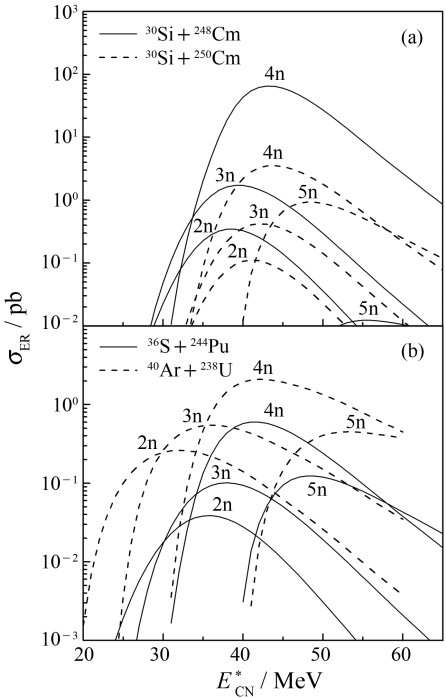

Fig. 13　Production cross sections of superheavy element Ds in the reactions $^{30}$Si+$^{248,250}$Cm, $^{36}$S+$^{244}$Pu and $^{40}$Ar+$^{238}$U.

## 3.2　Entrance channel effect in the production of SHN

The synthesis of heavy or superheavy nuclei through fusing two stable nuclei is inhibited by the so-called quasi-fission process. The entrance channel combinations of projectile and target will influence the fusion dynamics. The suppression of the evaporation residue cross sections for less fissile compound systems such as $^{216}$Ra and $^{220}$Th when reactions involved the projectiles heavier than $^{12}$C and $^{16}$O was observed experimentally in Ref.[66]. Fig. 14 is the excitation functions of 1n—5n channels in the reactions $^{34}$S+$^{238}$U, $^{64}$Fe+$^{208}$Pb and $^{136}$Xe+$^{136}$Xe which lead to the formation of the same compound nucleus $^{272}$Hs. The competition between the quasi-fission and the fusion process of the three systems leads to different trends of the evaporation channels. The 3n and 4n channels in the reaction $^{34}$S+$^{238}$U, 1n and 2n channels in the reaction $^{64}$Fe+$^{208}$Pb are favorable to produce the isotopes $^{268-269}$Hs and $^{270-271}$Hs. The larger transmission probabilities were found in the reactions $^{64}$Fe+$^{208}$Pb and $^{136}$Xe+$^{136}$Xe owing to the larger $Q$ values (absolute values). Smaller mass asymmetries of the two systems result in a decrease of the fusion probabilities[49]. Although the system $^{136}$Xe+$^{136}$Xe consists of two magic nuclei, the higher inner fusion barrier decreases the fusion probabilities and enhances the quasi-fission rate of the DNS, hence leads to the smaller cross sections of the Hs isotopes. The upper limit of the cross sections for evaporation residues $\sigma_{(1-3)n} \leqslant 4$ pb was reported in a recent experiment[67], which are much lower than the ones predicted by the fusion by diffusion model[68]. In the DNS model, the larger mass asymmetry favors the nucleon transfer from the light projectile to heavy target, and therefore enhances the fusion probability of two colliding nuclei.

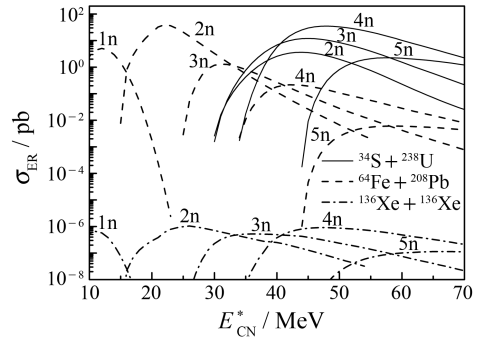

Fig. 14　Comparison of the calculated evaporation residue cross sections in 1n—5n channels for the reactions $^{34}$S+$^{238}$U, $^{64}$Fe+$^{208}$Pb and $^{136}$Xe+$^{136}$Xe.

### 3.3 Isotopic dependence of the production cross sections

Recent experimental data show that the production cross sections of the SHN depend on the isotopic combination of the target and projectile in the $^{48}$Ca induced fusion reactions. For example, the maximal cross section in the 3n channel is $3.7^{+3.6}_{-1.8}$ pb for the reaction $^{48}$Ca+$^{245}$Cm at the excitation energy 37.9 MeV; however, it is 1.2 pb for the reaction $^{48}$Ca+$^{248}$Cm although the later is a neutron-rich target[10, 69]. The isotopic dependence of the production cross sections were also observed and investigated in cold fusion reactions[24, 68]. In Fig. 15 we give the isotopic dependence of the production cross sections to synthesize the same SHN in the cold fusion and in the $^{48}$Ca induced reactions. It is shown that the isotopes $^{79}$Se based on $^{208}$Pb and $^{245, 247}$Cm in the 3n channels, $^{248}$Cm in the 4n channel, and $^{250}$Cm are suitable to produce the element $Z=116$. The $^{48}$Ca induced reactions give larger production cross sections than the cold fusion reactions. The corresponding excitation energies are also given in the figures. In the DNS model, the isotopic dependence of the production cross sections are mainly determined by both the fusion and survival probabilities. Of course, the transmission probability of two colliding nuclei can also be affected since the isotopes have initial quadrupole deformations. When the neutron number of the target increases, the DNS gets more asymmetrical and the fusion probability increases if the DNS does not consist of more stable nuclei (such as magic nuclei) because of a smaller inner fusion barrier. A smaller neutron separation energy and a larger shell correction lead to a larger survival probability. The compound nucleus with closed neutron shells has a larger shell correction energy and a larger neutron separation energy. Using neutron-rich actinide target has larger fusion and survival probabilities due to the larger asymmetric initial combinations and smaller neutron separation energies. But such actinide isotopes are usually unstable with smaller half-lives. With the establishment of the high intensity radioactive-beam facilities, the neutron-rich SHN may be synthesized experimentally, which approaches the island of sta-bility.

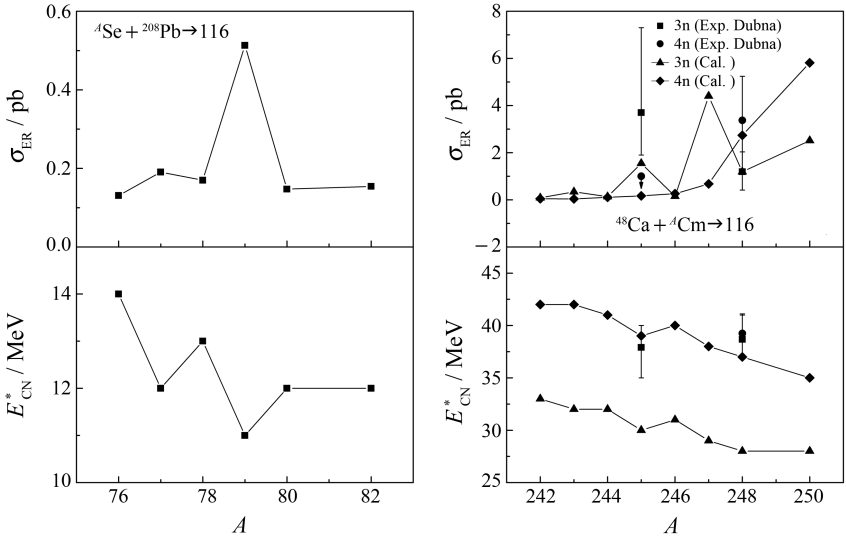

Fig. 15 Isotopic dependence of the calculated maximal production cross sections in cold fusion reactions and in $^{48}$Ca induced reactions leading to the synthesis of superheavy elements $Z=116$, and compared with the experimental data.

## 3.4 Production of neutron-rich heavy isotopes in low-energy transfer reactions

The cross sections of the heavy fragments in strongly damped collisions between very heavy nuclei were found to decrease very rapidly with increasing the atomic number[3-4]. Calculations by Zagrebaev and Greiner with a model based on multi-dimensional Langevin equations[21] showed that the production of the survived heavy fragments with the charged number $Z>106$ is rare because of the very small cross sections at the level of 1 pb and even below 1 pb. However, neutron-rich isotopes of Fm and Md were produced at the larger cross section of 0.1 $\mu$b. The evolution of the composite system in the damped collisions is mainly influenced by the incident energy and the collision orientation. Recently, the TDHF approach[29] and ImQMD model[28] were also used to investigate the dynamics in collisions of $^{238}$U+$^{238}$U.

The dynamics of the damped collisions of two very heavy nuclei was also investigated by the DNS model[70]. The cross sections of the primary fragments ($Z_1$, $N_1$) after the DNS reaches the relaxation balance are calculated as follows:

$$\sigma_{pr}(Z_1, N_1) = \frac{\pi \hbar^2}{2\mu E_{c.m.}} \times \sum_{J=0}^{J_{max}} (2J+1) P(Z_1, N_1, \tau_{int}). \quad (48)$$

The interaction time $\tau_{int}$ in the dissipation process of two colliding partners is dependent on the incident energy $E_{c.m.}$ in the center-of-mass (c.m.) frame and the angular momentum $J$, which has the value of few $10^{-20}$ s. The survived fragments are the decay products of the primary fragments after emitting particles and $\gamma$ rays in competition with fission. The cross sections of the survived fragments are given by

$$\sigma_{sur}(Z_1, N_1) = \frac{\pi \hbar^2}{2\mu E_{c.m.}} \sum_{J=0}^{J_{max}} (2J+1) \times P(Z_1, N_1, E_1, \tau_{int}) W_{sur}(E_1, xn, J), \quad (49)$$

where $E_1$ is the excitation energy of the fragment ($Z_1$, $N_1$). The maximal angular momentum is taken as $J_{max}=200$ that includes all partial waves in which the transfer reactions may take place. Shown in Fig. 16 is the calculated cross sections of the primary and survived fragments as functions of the charged numbers and mass numbers at the incident energy $E_{c.m.}=800$ MeV, respectively. In the damped collisions, the primary fragments result from a number of nucleon transfer in the relaxation process of the colliding partners. The giant composite system retains a very short time of several tens $10^{-22}$ s due to the strong Coulomb repulsion. Calculations from the TDHF method showed that the collision time depended on the orientation of the colliding system[29]. The cross sections in the production of heavy target-like fragments ($Z>92$) decrease drastically with the atomic numbers of the fragments. Therefore, the mechanism of the low-energy transfer reactions in collisions of two very heavy nuclei is not suitable to synthesize superheavy nuclei ($Z>106$) because of the smaller cross sections at the level of 1 pb and even below 1 pb.

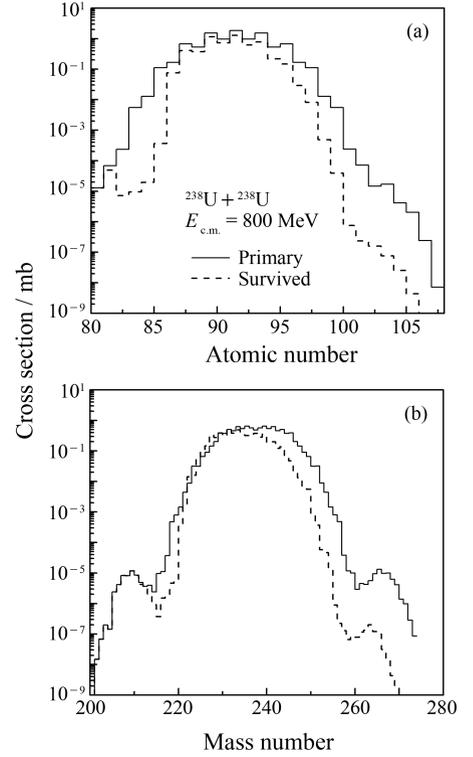

Fig. 16 Cross sections as functions of the charged and mass numbers of the primary and survived fragments at $E_{c.m.}=800$ MeV, respectively.

### 3.5 Synthesis of neutron-rich SHN with radioactive beams

The SHN with atomic number $Z=107—118$ synthesized by the cold fusion and the $^{48}$Ca induced reactions are all neutron-deficient in which the neutron number is smaller than the value at shell closure $N=184$. Further synthesis of heavier SHN with $Z>118$ becomes more and more difficult in the $^{48}$Ca induced reactions owing to the construction of heavy actinide target. Reactions with neutron-rich projectile bombarding actinide targets are possible to produce heavier SHN around $Z=120$ and 126, such as $^{64}$Ni+$^{238}$U, $^{70}$Zn+$^{232}$Th, $^{76}$Se+$^{238}$U and $^{86}$Kr+$^{232}$Th etc.. Radioactive ion beam of high intensity with the high $N/Z$ ratio may reach the domain of the island of stability predicted by shell models. Shown in Fig. 17 is a comparison between stable and weakly bound neutron-rich nuclei induced reactions to synthesize superheavy element $Z=114$. The larger cross section for the neutron-rich nuclei $^{56}$Ti is clear due to the larger survival probability resulting from the smaller neutron separation energy of compound nucleus. With the establishment of new-generation radioactive ion beam facility in the world, the construction is to be possible in the near future.

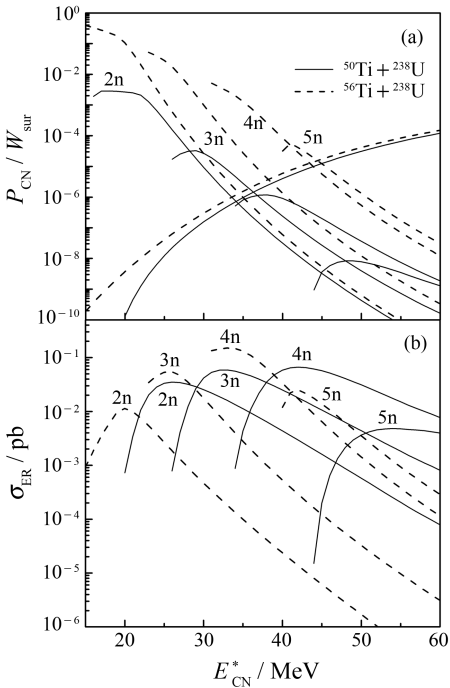

Fig. 17 Comparison of the fusion, survival probabilities and final production of SHN in the reactions $^{50}$Ti+$^{238}$U and $^{56}$Ti+$^{238}$U.

## 4 Conclusions

Recent progress of the theoretical models on describing the formation of superheavy nuclei is reviewed and discussed. Calculated results based on the DNS model are presented and compared with the available experimental data. Reactions based on $^{238}$U and new isotopes of element Ds with stable neutron-rich projectile bombarding actinide targets are discussed. Both of the reaction $Q$ value and the mass asymmetry in the entrance channel influence the final products of SHN. Sub-shell closure at $N=162$ enhances the production of neutron-rich heavy isotopes in low-energy transfer reactions of actinides. Synthesis of neutron-rich SHN with radioactive beams are discussed.

# 低能重离子碰撞产生超重核动力学机制*

冯兆庆[1]，靳根明，李君清

（中国科学院近代物理研究所，甘肃 兰州 730000）

**摘　要**：总结了描述重系统碰撞形成超重核的主要理论模型进展。基于两类反应机制，即熔合蒸发反应和大质量阻尼反应，对产生超重核的物理过程进行了讨论。分析了超重核形成过程中碰撞系统的俘获、复合核的形成及蒸发退激描述存在的问题。基于双核模型分析了合成冷熔合反应和 $^{48}$Ca 诱发全熔合反应之间的超重新核素的可能性。利用锕系核碰撞的转移反应产生子壳 $N=162$ 附件丰中子重核的可行性，研究了壳效应对丰中子核素产生的影响。进一步讨论了将来基于丰中子强流放射性束合成超重核的可行性。

**关 键 词**：超重核；熔合蒸发反应；阻尼碰撞；熔合动力学